\documentclass[12pt,preprint]{aastex}
\begin{document}

\title{The Giant X-Ray Flare of NGC~5905: Tidal Disruption of\\a Star, a 
Brown Dwarf, or a Planet?}

\author{Li-Xin Li\altaffilmark{a,1},\, Ramesh Narayan\altaffilmark{a},\,
and Kristen Menou\altaffilmark{b,1}}
\affil{$^{a}$Harvard-Smithsonian Center for Astrophysics, Cambridge, MA 02138, USA}
\email{lli,rnarayan@cfa.harvard.edu}
\affil{$^{b}$Princeton University Observatory, Princeton, NJ 08544, USA}
\email{kristen@astro.princeton.edu}

\altaffiltext{1}{Chandra Fellow}

\begin{abstract}
We model the 1990 giant X-ray flare of the quiescent galaxy NGC~5905
as the tidal disruption of a star by a supermassive black hole. From
the observed rapid decline of the luminosity, over a timescale of a
few years, we argue that the flare was powered by the fallback of
debris rather than subsequent accretion via a thin disk. The fallback
model allows constraints to be set on the black hole mass and the mass
of debris.  The latter must be very much less than a solar mass to
explain the very low luminosity of the flare.  The observations can be
explained either as the partial stripping of the outer layers of a
low-mass main sequence star or as the disruption of a brown dwarf or a
giant planet.  We find that the X-ray emission in the flare must have
originated within a small patch rather than over the entire torus of
circularized material surrounding the black hole.  We suggest that the 
patch corresponds to the ``bright spot'' where the stream of returning 
debris impacts the torus.  Interestingly, although the peak
luminosity of the flare was highly sub-Eddington, the peak flux from
the bright spot was close to the Eddington limit.  We speculate on the
implications of this result for observations of other flare events.
\end{abstract}


\section{Introduction}
\label{sec1}

Tidal disruption of a star by a supermassive black hole in a galactic
nucleus has been investigated by many authors.  Early seminal
investigations by \citet{hil75,lac82,ree88,eva89} and \citet{can90}
were followed up with detailed analytical studies
\citep{koc94,kho96,loe97,ulm98,ulm99} and numerical simulations
\citep{kho93,fro94,mar96,die97,kim99,aya00,iva01}.

A tidal disruption event is expected to produce a luminous flare of
electromagnetic radiation in the UV to X-ray band.  \citet{kom01}
argued that a convincing detection of a tidal disruption event must
fulfill the following three criteria: (1) The event should be of short
duration (a ``flare''); (2) It should be very luminous (up to
$L_{\max} \sim 10^{45}$ erg~s$^{-1}$ at maximum); (3) It should reside 
in a galaxy that is otherwise non-active (in order to exclude an upward
fluctuation in the mass accretion rate of an active nucleus).

{\it ROSAT} observations have led to the discovery of four X-ray
flaring ``normal'' galaxies that fulfill the above criteria
[summarized in \citet{kom01}]: NGC~5905, RXJ1242-1119, RXJ1624+7554,
RXJ1420+5334; plus a possible fifth candidate: RXJ1331-3243. All these
galaxies have exhibited large X-ray flares (peak luminosity up to
$\sim 10^{44}$ erg~s$^{-1}$), corresponding to a significant amplitude
of variability (up to a factor $\sim 200$), an ultra-soft X-ray
spectrum ($k_B T_{\rm bb} \approx 0.04-0.1$ keV, where $T_{\rm bb}$ is
the temperature derived from a black-body fit to the spectral data),
and no signs of Seyfert-like activity in the optical. These flares are
promising candidates for tidal disruption events \citep{kom01}. A UV
flare has also been detected at the center of a mildly active
elliptical galaxy, NGC~4552, with a bolometric luminosity $\sim
10^{39}$ erg~s$^{-1}$.  Since the luminosity is several orders of
magnitude less than would be expected from complete tidal disruption
of a star, the event has been explained as the tidal stripping of a
stellar atmosphere \citep{ren95}.

Among the five galaxies listed above, long-term follow-up observations
exist only for NGC~5905 \citep{kom99}. In this paper, we model the
light curve of NGC~5905 assuming that the flare was produced by the
tidal disruption of a star by a supermassive black hole.  We compare
theoretical predictions with the observational data and constrain the
properties of the progenitor star (e.g., its mass and radius) and
estimate how much mass was accreted by the black hole.

\section{The Light Curve of the Flare in NGC~5905}
\label{sec2}

\begin{figure}
\epsscale{0.9}
\plotone{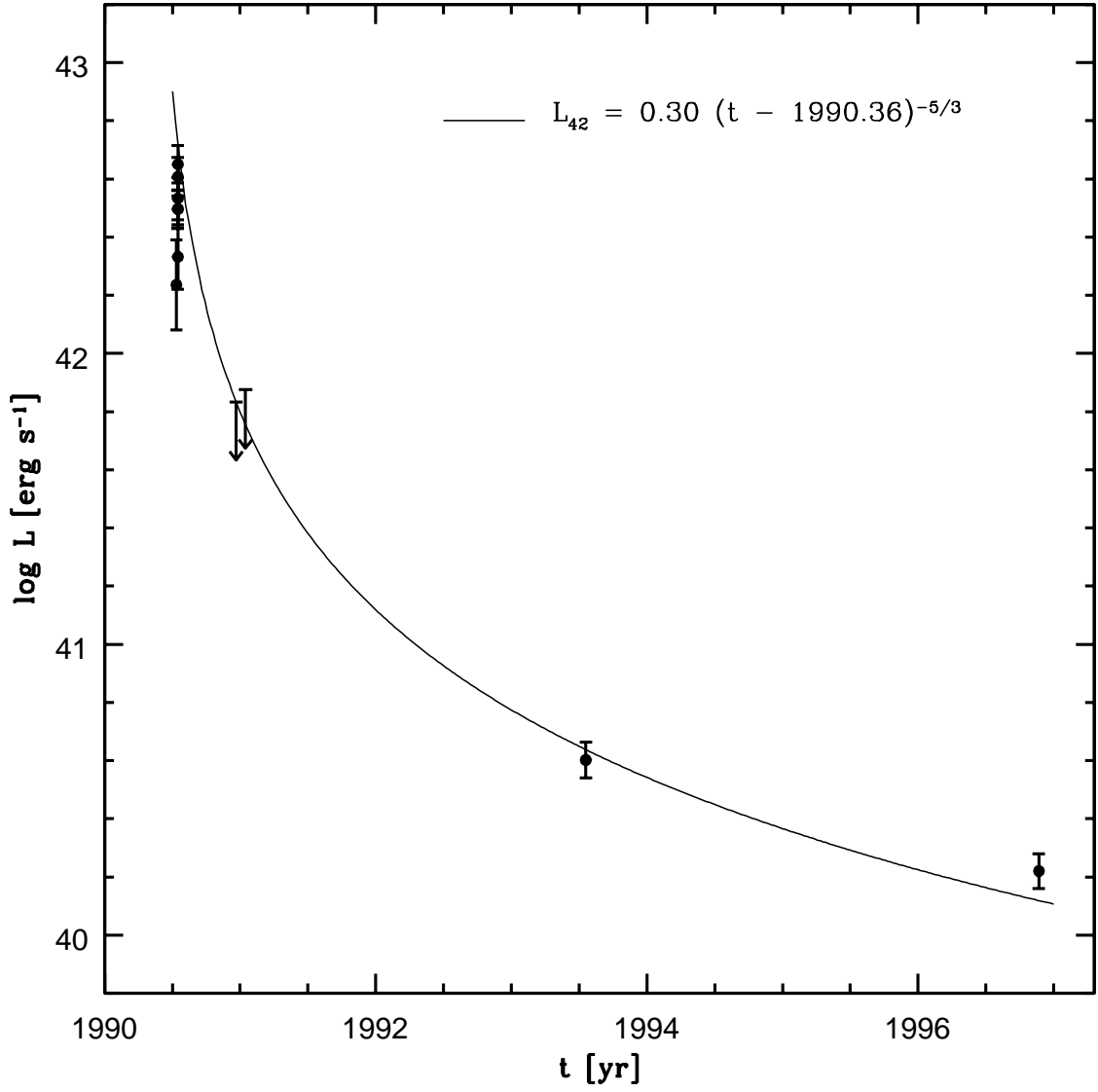}
\caption{The X-ray light curve of the flare in NGC~5905 (see Table 1
for the data), fitted to a model of the form $(t-t_D)^{-5/3}$, as
expected for fallback of material after a tidal disruption.
\label{fig1}}
\end{figure}

A giant X-ray flare in NGC~5905 was detected by {\it ROSAT} in July
1990 \citep{bad96}. The observed peak X-ray luminosity exceeded
$10^{42}$ erg~s$^{-1}$, and the X-ray spectrum was very soft (photon
index $\Gamma_X\approx -4$, or a very cool black-body with $k_BT_{\rm
bb}\approx0.06$ keV). Follow-up observations were made in 1990 and
1993 with the PSPC, and in 1996 with the HRI \citep{kom99}. The
observational data are listed in Table \ref{tab1}. Assuming a Hubble
constant $H_0 = 50$ km s$^{-1}$ Mpc$^{-1}$, the count rates may be
converted to luminosities, which are plotted with error bars in Figure
\ref{fig1}. The two downward arrows are upper limits. The light curve
after the initial burst gradually declines with time. During the X-ray 
outburst as well as at later times, the optical brightness is constant 
to within observational errors ($\sim 0.2^m$).

Fitting the late decline of the NGC~5905 light curve (the four last
data points in Table~\ref{tab1}) with a $(t - t_D)^{-5/3}$ law that is
predicted by the fallback model of tidal disruption (see \S\ref{sec3}), 
and treating the largest luminosity observed at $t=1990.54$ yr, viz.
$L_X=4.47\times10^{42} ~{\rm erg\,s^{-1}}$, as a lower limit (since
this was still in the rising part of the light curve), we obtain
\begin{eqnarray}
        L_X = (0.30\pm 0.03) \times 10^{42}\, {\rm erg\, s^{-1}}\,
              \left[\frac{t - (1990.36\pm 0.02)\, {\rm yr}}{1\,
              {\rm yr}}\right]^{-5/3} \,,
        \label{lum2}
\end{eqnarray}
where the time $t$ is in units of years, and $t_D = 1990.36\pm 0.02$ yr
is the moment when tidal disruption happened.

During the first epoch of observations at time $t=1990.54$, the X-ray
flux rose steadily (see Table 1), strongly suggesting that this epoch
corresponded to the onset of the flare.  For a start time of $t_1
\approx 1990.54$ yr, the total X-ray energy released during the entire
outburst event is
\begin{eqnarray}
        \Delta E_X = \int_{t_1}^\infty L_X (t) dt \approx (4.5
                     \pm 0.9) \times 10^{49}\, {\rm erg}\,.
        \label{ex}
\end{eqnarray}

A black-body fit to the X-ray spectrum during the outburst gives a
temperature $k_BT_{\rm bb} \approx 0.06$ keV, and an intrinsic ($0.1 -
2.4$ keV) mean luminosity $L_X \approx 3 \times 10^{42}$
erg~s$^{-1}$. Integrating the black-body spectrum over all energies
(from zero to infinity), the bolometric luminosity is $L \approx 1.1
L_X$. Thus, if the spectrum does indeed have a black-body form, the
bolometric correction is not very significant.  Then, the total mass
associated with the released energy may be estimated to be
\begin{eqnarray}
        \Delta M = \frac{\Delta E}{\epsilon c^2} \approx
                   \frac{\Delta E_X}{\epsilon c^2} \approx 
                   (2.5 \pm 0.5) \times 10^{-4}\, M_{\odot}\,
                   \left(\frac{\epsilon}{0.1}\right)^{-1}\,,
        \label{mass}
\end{eqnarray}
where $\epsilon$ is the efficiency of converting mass to radiated
energy.  This mass estimate is similar to that for the UV flare in
NGC~4552 \citep{ren95}.  Note that our estimate of $\Delta E_X$
depends only weakly on the assumed power law index $5/3$ in equation
(\ref{lum2}).  This is because the observed light curve extends over a
long enough time to capture most of the emission in the flare.  The
estimate of $\Delta M$ suffers from the uncertainty in the bolometric
correction (which we have taken to be unity).

In models of thin accretion disks around black holes, the color
temperature $T_{\rm bb}$ of the emitted radiation is generally higher
than the effective temperature $T_{\rm eff}$, defined by flux
$F=\sigma T_{\rm eff}^4$.  The deviation of $T_{\rm bb}$ from $T_{\rm
eff}$ is caused by electron scattering and Comptonization (Ross,
Fabian \& Mineshige 1992; Shimura \& Takahara 1993, 1995a, 1995b).
Writing $T_{\rm bb} = f_c T_{\rm eff}$, the factor $f_c$ can be
estimated from detailed self-consistent models of the radiative
transfer in the disk.  It is found that the value of $f_c$ depends on
both the mass of the black hole (compare Shimura \& Takahara 1993 and
1995b) and the mass accretion rate.  Shimura \& Takahara (1993)
present results for a $10^8M_\odot$ black hole emitting at a radius of
5 Schwarzschild radii, which is similar to the conditions we expect in
our problem.  Depending on the accretion rate, they obtain values of
$f_c$ in the range $1.3-3.2$ (see their Fig. 4 and the associated
discussion), with the highest value of $f_c$ being obtained for
near-Eddington rate accretion. In what follows, we retain the factor
$f_c$ in the equations, and we substitute a value of 3 whenever we
require a numerical estimate.

Including the factor $f_c$ as defined above, we estimate the radius of
the X-ray emitting region of the flare in NGC 5905 to be
\begin{eqnarray}
        R_X = \left(\frac{f_c^4 L_X}{\pi\sigma T_{\rm bb}^4}\right)^{1/2}
              \approx 2.4 \times 10^{12}\,{\rm cm}\,
              \left(\frac{f_c}{3}\right)^2\,,
              \qquad f_c \equiv {T_{\rm bb}\over T_{\rm eff}}\,,
        \label{rx}
\end{eqnarray}
where we have used $L_X \approx 3 \times 10^{42}$ erg~s$^{-1}$ for the
mean luminosity during the peak of the outburst \citep{kom01}. The
radius we derive is surprisingly small; even after setting $f_c=3$, it
is barely equal to the Schwarzschild radius of a $10^7M_\odot$ black
hole, whereas an accretion disk is likely to have an effective
radiating area $10-100$ times larger.  We return to this point in \S5.

\section{Tidal Disruption of a Star by a Supermassive Black Hole}
\label{sec3}

The tidal disruption radius of a star in the vicinity of a
Schwarzschild black hole is 
\begin{eqnarray}
	r_T = \mu R_\star \left({M_H\over M_\star}\right)^{1/3} \;,
\label{rT}
\end{eqnarray}
where $M_H$ is the mass of the black hole, $M_\star$ is the mass of
the star, $R_\star$ is the radius of the star, and $\mu$ is a
dimensionless coefficient.  The exact value of $\mu$ is uncertain,
though physically we expect it to be of order unity.  We assume
$\mu=1$ in our modeling.

The ratio of $r_T$ to $r_H$, where $r_H = 2 GM_H/c^2$ is the
Schwarzschild radius of the black hole, is
\begin{eqnarray}
        \frac{r_T}{r_H} = 5.08 \left(\frac{M_H}{10^7 M_\odot}
              \right)^{-2/3} \left(\frac{M_\star}{M_\odot}\right)^{-1/3}
              \left(\frac{R_\star}{R_\odot}\right) \,.
        \label{rth}
\end{eqnarray}
For a solar-type star with $M_\star = M_\odot$ and $R_\star =
R_\odot$, we have $r_T < r_H$ when $M_H > 1.1 \times 10^8\, M_\odot$.
In such a case the star will be swallowed whole by the black hole,
without a tidal disruption event \citep{hil75}.

After a star is tidally disrupted, there are two distinct stages in
the dynamical evolution of the debris, each with its own time
development and luminosity:

(1) The fallback stage \citep{ree88}: A fraction of the material in
the disrupted star remains gravitationally bound to the black hole and
falls back to the pericenter, giving rise to a mass accretion rate
(and a luminosity) evolving with time as $\sim t^{-5/3}$
\citep{phi89}. The returning material shocks and then circularizes to
form an orbiting torus around the black hole at a radius $r_c = 2 r_P$
(by angular momentum conservation), where $r_P$ is the pericentric
radius.  The total energy radiated during the fallback is given by
$GM_H\Delta M/2r_c$, where $\Delta M$ is the fallback mass.  This
energy can be less than the energy released in the subsequent viscous
accretion stage discussed below.  Nevertheless, since the fallback
stage is relatively short-lived, it dominates the early luminosity of
a disruption event.

(2) The viscous accretion stage \citep{can90}: The torus formed from
returning debris gradually spreads inward and outward from $r = r_c$
under the action of viscosity, and gives rise to a mass accretion
rate (and a luminosity) evolving with time approximately as $\sim
t^{-1.2}$. The energy radiated during this stage is given by $GM_{\rm 
H}\Delta
M[(1/2r_{ms}) -(1/2r_c)]$, where $r_{ms}$ is the radius of the last
stable circular orbit ($=3$ Schwarzschild radii for a non-spinning
black hole).  The accretion occurs on a viscous time scale, which is
typically very long compared to the fallback time (Cannizo et
al. 1990; Ulmer 1999; Appendix A of the present paper).  Therefore,
this stage is expected to last a long time, up to hundreds of years,
depending on the magnitude of the viscosity in the disk, and the
luminosity is significantly lower than that associated with the
fallback stage (Appendix \ref{appa}).

In the case of NGC~5905, the X-ray luminosity was seen to rise rapidly
during the first epoch of observations at $t = 1990.54$ yr, indicating
that the flare probably began around that time (see Table 1 and
Fig.~\ref{fig1}). The luminosity then dropped by more than a factor of
$100$ between $t = 1990.54$ yr and $ t = 1996.89$ yr.  Such a rapid
decline indicates that the observed outburst of NGC~5905 could not
have been due to the accretion phase, but must have corresponded to
the fallback stage (if it indeed was a tidal disruption
event). Therefore, in this paper we assume that the flare in NGC~5905
corresponds to the fallback stage of tidal disruption.

Let us assume that the center of the star follows a nearly parabolic
orbit (i.e., the binding energy of the star to the black hole is close
to zero), and for definiteness let us assume that the pericenter of
the orbit is at $r_P = r_T$, i.e. that the parameter $\eta$ defined
below in equation (\ref{omgs}) is equal to unity.  Since the specific
energy at the center of the star is $E_c = v_P^2/2 - G M_H/r_T = 0$,
the orbital velocity of the star at pericenter is $v_P = (2 G M_H /
r_T)^{1/2} = c\, (r_H / r_T)^{1/2}$, and the spread in the specific
energy of the disrupted debris, $2\Delta E$, is governed by the
variation of the black hole gravitational potential across the star,
and the spin-up of the star as a result of the tidal interaction
\citep{ree88}. If $M_H \gg M_\star$, we have
\begin{eqnarray}
        \Delta E = k \frac{G M_\star}{R_\star} \left(\frac{M_H}{M_\star}
                    \right)^{1/3} \,,
        \label{dele}
\end{eqnarray}
where $k$ depends on the spin-up state of the star. If the star is spun up 
to the break-up spin angular velocity, we have $k \approx 3$. On the other 
hand, if the spin-up effect is negligible, then we have $k \approx 1$ 
\citep{ree88,lac82,eva89,aya00}. Since the spin-up velocity is always
$\ll v_P$, the energy used to spin up the star is much less than the 
orbital kinetic energy of the star, and the stellar orbit remains parabolic 
to a very good approximation. 

The spin-up of a star via tidal interaction is a complex process.  In
linear perturbation theory, the spin-up angular velocity is given by
\citep{pre77,ale01a,ale01b}
\begin{eqnarray}
        \omega_s \approx \frac{T_2 (\eta)}{2 \alpha\eta^2}\, \omega_P\,,
        \qquad \eta \equiv \left({r_P\over r_T}\right)^{3/2}\,,
        \label{omgs}
\end{eqnarray}
where $\eta \approx1$ in our problem, $\omega_P 
\equiv v_P / r_P$ is the orbital angular velocity of the star at the 
pericenter, $\alpha$ is the stellar momentum of inertia in units of $M_\star
R_\star^2$, and $T_2$ is the second tidal coupling coefficient which
is a function of $\eta$. In equation (\ref{omgs}) we have assumed that
$M_H \gg M_\star$. For a main sequence star of mass $0.76 M_\odot$ and
radius $0.75 R_\odot$, \citet{ale01a} give $\alpha \approx 0.07$ and
$T_2(1) \approx 0.06$, which implies that $\omega_s \approx 0.43
\omega_P$ if $\eta = 1$.  For an $n = 1.5$ polytrope star, they give 
$\alpha\approx 0.21$ and $T_2(1) \approx 0.36$, which corresponds to 
$\omega_s \approx 0.86 \omega_P$.  Furthermore, they show that numerical 
simulations that include non-linear effects lead to a larger energy 
transfer from the orbit to the star and a larger spin-up than predicted 
by linear theory.  For these reasons, we feel that it is likely a tidally
disrupted star will be spun up to nearly the
break-up limit, with $\omega_s \approx
\omega_P$ and $k \approx 3$. We therefore assume $k = 3$ for most of
our analysis; however, for completeness we also present a few results
for the non-spinning case $k=1$.

Part of the disrupted star has positive energy and escapes from the
system, and part remains bound.  The bound material follows a highly
eccentric orbit and returns to the central black hole after completing
one orbit. The flare begins when the most bound material, with
specific energy $E_{\min} = - \Delta E$, returns.  The time since
disruption for this to happen is given by
\begin{eqnarray}
        \Delta t_1 = 2\pi G M_H (2 \Delta E)^{-3/2} \approx
               0.068\,{\rm yr}\, \left(\frac{M_H}{10^7 M_\odot}
              \right)^{1/2} \left(\frac{M_\star}{M_\odot}\right)^{-1}
              \left(\frac{R_\star}{R_\odot}\right)^{3/2} \,,
        \label{t1}
\end{eqnarray}
where we have used $r_P = r_T$ and $k = 3$ \footnote{If the star is on a
parabolic orbit with $r_P < r_T$, then the spin-up effect will be less
important and we will have a smaller $k$ (but $k$ is still $\ga 1$), 
which will give rise to a $\Delta t_1$ that is somewhat larger than that 
given by eq. (\ref{t1}).}.  Material that is less
bound takes a progressively longer time to return, and the overall
light curve is determined by the distribution of debris mass as a
function of binding energy.

\citet{ree88} assumed that the debris is uniformly distributed in mass
between $-\Delta E$ and $+ \Delta E$.  Numerical simulations 
\citep{eva89,aya00} have shown that this is a reasonable approximation. 
The bound material then returns to pericenter at the rate 
\citep{phi89,eva89}
\begin{eqnarray}
        \dot{M} = \frac{2\Delta M}{3 \Delta t_1} \left(\frac{t - t_D}{\Delta 
                  t_1}\right)^{-5/3}
                = A \left(\frac{t - t_D}{1 {\rm yr}}\right)^{-5/3} \,
                \qquad \Delta M \equiv f M_\star\,,
        \label{mdot}
\end{eqnarray}
where $t_D$ is the time of the initial tidal disruption, 
$\Delta M$ is the actual mass that falls back to pericenter, which is a
fraction $f$ of the original mass of the star, and
\begin{eqnarray}
        A \approx 7.0 \times 10^{23} \,{\rm g~s^{-1}} \,
            \left(\frac{f}{0.1}\right) \left(\frac{M_H}{10^7 M_\odot}
            \right)^{1/3} \left(\frac{M_\star}{M_\odot}\right)^{1/3}
            \left(\frac{R_\star}{R_\odot}\right) \,.
        \label{mdot1}
\end{eqnarray}

In Rees' original
model, the whole star is disrupted and half the debris is on bound
orbits, so that $f = 0.5$. However, a recent numerical simulation
shows that not all the returning material is captured by the black
hole: about $75\%$ of the returning mass becomes unbound following the
large compression it experiences on the way back \citep{aya00}. This
gives rise to a smaller $f\approx 0.1$. Another possibility giving
rise to a small $f$ is that the star is only partially disrupted: its
envelope could be stripped by the black hole, leaving most of its core
nearly intact (Renzini et al. 1995; see also \S5 below).  Here we
treat $f$ as a free parameter.

The gravitational potential energy available from fallback is
determined by the difference between the specific binding energy of
the circularization orbit at $r = 2 r_P = 2r_T$ and the specific
binding energy of the incoming material. Since $M_H \gg M_\star$, all
the bound debris is on highly eccentric orbits with a specific binding
energy much smaller than the binding energy of the final circular
orbit. Thus, assuming that the fallback material radiates the energy
release promptly, the radiation efficiency $\epsilon$ is independent
of time during fallback:
\begin{eqnarray}
        \epsilon \approx \frac{G M_\star}{4 R_\star c^2}\left(\frac{M_H}
                         {M_\star}\right)^{2/3} \approx 0.025
                         \left(\frac{M_H}{10^7 M_\odot}\right)^{2/3}
                         \left(\frac{M_\star}{M_\odot}\right)^{1/3}
                         \left(\frac{R_\star}{R_\odot}\right)^{-1}\,.
        \label{ep}
\end{eqnarray}
The luminosity of the fallback process is then given by
\begin{eqnarray}
        L = \epsilon \dot{M} c^2 \approx 1.55 \times 10^{43}\, 
                {\rm erg~s^{-1}}\,\left(\frac{f}{0.1}\right)
                \left(\frac{M_H}{10^7 M_\odot}\right)
                \left(\frac{M_\star}{M_\odot}\right)^{2/3}\,
                \left(\frac{t-t_D}{1 {\rm yr}}\right)^{-5/3} \,.
            \label{rees_lum}
\end{eqnarray}
The luminosity peaks at $t = t_D + \Delta t_1$, i.e. when the most
bound debris falls back to the pericenter, so we have
\begin{eqnarray}
        L_{\rm peak} \approx 1.36 \times 10^{45}\, {\rm erg~s^{-1}}\,
                \left(\frac{f}{0.1}\right)
                \left(\frac{M_H}{10^7 M_\odot}\right)^{1/6}
                \left(\frac{M_\star}{M_\odot}\right)^{7/3}
                \left(\frac{R_\star}{R_\odot}\right)^{-5/2}\,.
            \label{lum_peak}
\end{eqnarray}

\section{Comparison with the Flare in NGC~5905}
\label{sec4}

In \S\ref{sec2} we have fitted the decline of the NGC~5905 light curve
to a $(t - t_D)^{-5/3}$ law, and obtained the result given in equation
(\ref{lum2}) with a disruption time of $t_D=1990.36$ yr.  The fit is
acceptable (see Fig. \ref{fig1}), and the time at which the light
curve reached its peak is fairly well constrained.  Since the initial
measurements at $t \approx 1990.54$ yr showed flux rising with time,
the maximum in the light curve was clearly later than this time.  At
the same time, the maximum could not have been later than 1990.56 yr,
since the fitted light curve in (\ref{lum2}) falls below $L \approx
4.47 \times 10^{42}$ erg~s$^{-1}$ at that time.  Thus, the light curve 
fit gives $\Delta t_1$ within the range 0.18 to 0.20 yr, where $\Delta
t_1$ is the time lag between the peak of the light curve and 
the moment of tidal disruption.  However, there is an error of $\pm
0.03$ yr on each limit.  Therefore, we use conservative bounds on 
$\Delta t_1$
\begin{eqnarray}
        0.15\, {\rm yr} < \Delta t_1 < 0.23\, {\rm yr} \,.
        \label{t1l}
\end{eqnarray}

>From the discussion in \S\ref{sec2}, the mass involved in the outburst
of NGC~5905 is only $\sim 10^{-4}M_\odot$. We therefore consider the
following two possibilities for the candidate stellar object disrupted
in NGC~5905:

\noindent
(1) A low mass star (LMS), whose radius and mass are related by
$R_\star = R_\odot (M_\star /M_\odot)$ $(0.08 M_\odot < M_\star < 1
M_\odot$).

\noindent
(2) A substellar objects (SSOs, including brown dwarfs and planets),
whose radius and mass are related by $R_\star = 0.06 R_\odot (M_\star
/ M_\odot)^{-1/8}$ ($0.01 M_\odot < M_\star < 0.08 M_\odot$)
\citep{cha00}\footnote{\citet{cha00} state that their fit is good down
to $M_\star=0.01M_\odot$, but a comparison with their plot shows that
the fit is acceptable even for lower masses. We apply their scaling
down to $M_\star\sim0.006M_\odot$.}.

\begin{figure}
\epsscale{0.83}
\plotone{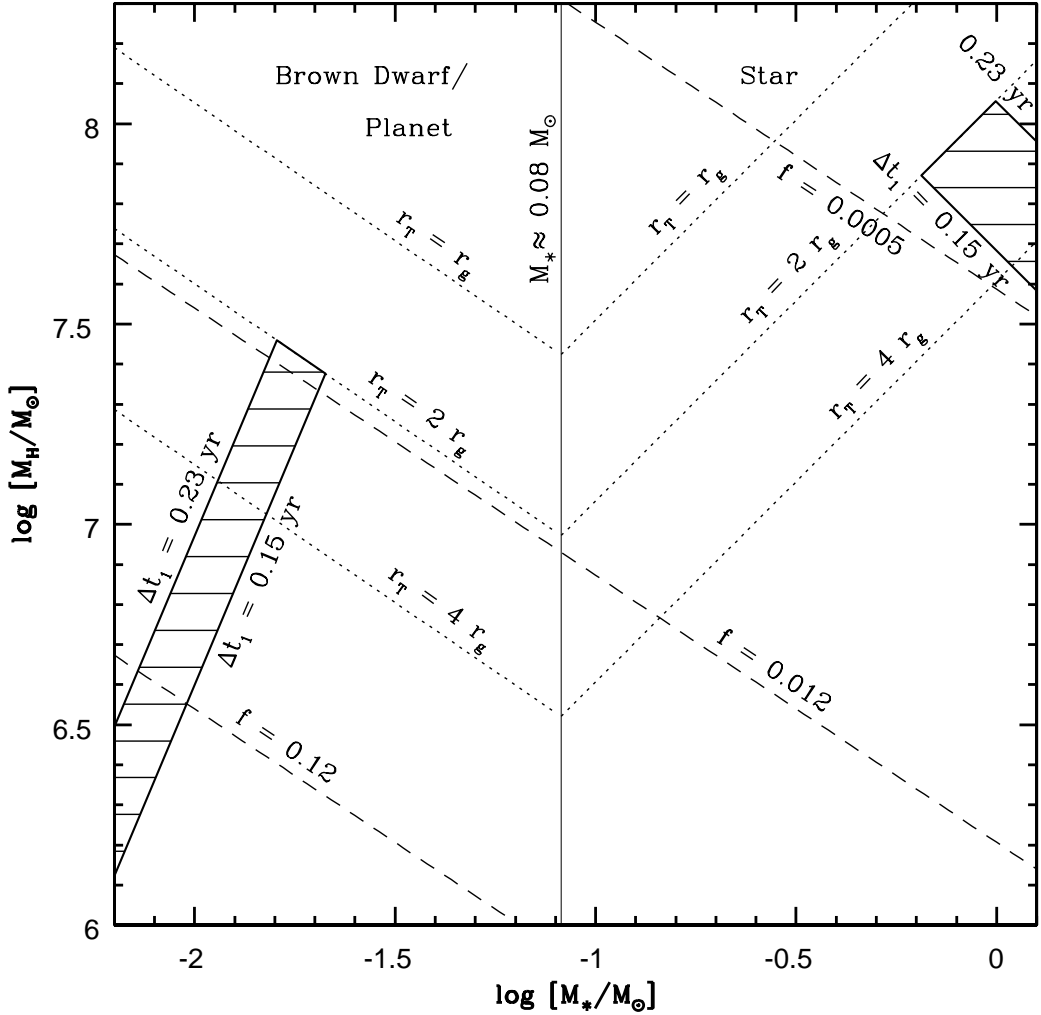}
\caption{Restrictions on the mass of the black hole, $M_H$, and the
mass of the disrupted stellar object, $M_\star$, derived from the
X-ray flare of NGC~5905.  The shaded areas show the allowed regions.
It has been assumed that the star is spun up to breakup at pericenter
($k=3$).  The region to the left of the thin vertical line corresponds
to brown dwarfs/planets, and the region to the right to low mass
stars. The limits used are (1) $0.15\,{\rm yr} < \Delta t_1 < 0.23\,
{\rm yr}$, where $\Delta t_1$ is the time for the most bound disrupted
material to fall back to pericenter; and by (2) $r_T > 2 r_g$, where
$r_T$ is the tidal disruption radius (assumed to be equal to the
pericentric radius) and $r_g \equiv G M_H /c^2$.  For comparison,
lines corresponding to $r_T = r_g$ and $r_T = 4 r_g$ are also shown.
The dashed lines correspond to equation (\ref{ml}) for various values
of the fraction $f$ of the mass of the disrupted star that falls back.
\label{fig2}}
\end{figure}

Substituting these two radius-mass relations into equation (\ref{t1})
and noting the limits on $\Delta t_1$ in equation (\ref{t1l}), we find
that the black hole mass in NGC 5905 and the mass of the disrupted
star are restricted to the two relatively narrow shaded bands shown in
Figure \ref{fig2}.  The figure makes use of an additional constraint,
namely a lower limit on the tidal radius $r_T$.  If the black hole is
not rotating, then the smallest distance to which a star on a
parabolic orbit can approach and yet not be captured by the black hole
is $r=4r_g$, where $r_g\equiv GM_H/c^2=r_H/2$. If we have a maximally
rotating black hole and if the angular momentum of the stellar orbit
is parallel to the spin of the black hole, then the closest approach
possible is $r=r_g$.  The real situation is probably in between these
two limits, so we use a fiducial limiting radius of $r=2r_g$.  Using
equation (\ref{rth}), we have calculated relations connecting $M_H$
and $M_\star$ for all three limiting radii and plotted them as dotted
lines in Figure~\ref{fig2}.

The mass of the black hole in NGC 5905 is not very well constrained,
but it is believed to be in the range $M_H \sim 10^7 - 10^8 M_\odot$
\citep{kom01}, based on the correlation between bulge blue luminosity
and black hole mass for spiral galaxies \citep{sal00} and on the
correlation between bulge velocity dispersion and black hole mass for
ellipticals and spirals \citep{geb00,mer01}.  This information is 
sufficient to constrain the mass of the disrupted star fairly tightly: 
the mass must be either in the range $0.01-0.02 M_\odot$ in the case of 
a SSO or $0.6-1M_\odot$ in the case of a LMS.

The result, however, depends sensitively on the choice we make for the
tidal spin-up parameter $k$ discussed in \S2.  Figure \ref{fig2} has
been calculated for $k=3$, a reasonable and possibly likely value, but
in principle $k$ could be as small as 1.  Figure \ref{fig3} shows how
the mass constraints change when we use the latter value.  We see that
the allowed range of models is limited to somewhat lower black hole
masses, unless the black hole spins very rapidly and allows tidal
disruptions down to $r_T\sim r_g$.  We do not discuss the $k=1$ case
further.

>From equations (\ref{rees_lum}) and (\ref{lum2}) we infer that
\begin{eqnarray}
        \frac{M_H}{10^7 M_\odot}\, \left(\frac{M_\star}
                {M_\odot}\right)^{2/3} = \frac{0.0019}{f}\,,
        \label{ml}
\end{eqnarray}
where $f$ is the fraction of the mass of the star that returns as
fallback debris.  The dashed lines in Figure \ref{fig2} correspond to
$f = 0.0005$, $0.012$, and $0.12$ respectively. We see that, for the
case of a LMS, $f$ must be $< 0.0005$, which means that an unusually
small fraction of the star must have participated in the fallback. For
the case of a SSO, somewhat larger values of $f$ are obtained (but
still rather small).

In Figures \ref{fig4} and \ref{fig5} we plot the luminosities given by
equation (\ref{rees_lum}) for different values of $M_H$, $M_\star$,
and $f$, where the disrupted object is assumed to be a LMS
(Fig.~\ref{fig4}) and a SSO (Fig.~\ref{fig5}), respectively, and the
disruption time is taken to be $t_D = 1990.36$ yr.  We have chosen
values of $(M_H, M_\star)$ from the shaded regions in Figure
\ref{fig2} so that equation (\ref{t1l}) and the condition $r_T > 2
r_g$ are satisfied, and chosen $f$ such that $L_{\rm peak} \ga 4.47
\times 10^{42}$ erg~s$^{-1}$ where $L_{\rm peak} \equiv L(t = t_D +
\Delta t_1)$ is the peak luminosity.

We confirm that, in order for the models to fit the observational
data, a very small $f$ ($<0.0003$) is required for the case of a LMS
(Fig.~\ref{fig4}). For the case of a SSO (Fig.~\ref{fig5}) a fairly
small $f$ ($\la 0.04$) is again required if $M_H \ga 2\times 10^7 M_\odot$.
A largish $f$ ($> 0.1$) can be obtained only if the black hole has a
very small mass: $M_H < 4 \times 10^6 M_\odot$.

In Figures \ref{fig4} and \ref{fig5} we also indicate the ratio of the
peak flux, $F_{\rm peak}$, in the fitted light curve to the Eddington
flux, $F_{\rm Edd}$, where $F_{\rm peak} \equiv L_{\rm peak}/(\pi
R_X^2)$ with $R_X$ given by equation (\ref{rx}), and $F_{\rm Edd}
\equiv L_{\rm Edd}/(4 \pi r_T^2)$, with $L_{\rm Edd} = 1.3 \times
10^{45}\,{\rm erg~s^{-1}}\, (M_H/10^7 M_\odot)$ and $r_T$ given by
equation (\ref{rth}).  We see that the ratio $F_{\rm peak}/F_{\rm
Edd}$ is significantly larger than unity if the color temperature
factor $f_c\sim1$, and it is not much less than unity even if
$f_c\sim3$ (the likely value).  We discuss the implications of this
result in the next section.

\section{Summary and Discussion}
\label{sec5}

The X-ray light curve of the flare in NGC~5905 agrees with the
$(t-t_D)^{-5/3}$ dependence predicted for fallback of debris after
tidal disruption of a star by a supermassive black hole
(Fig. \ref{fig1}).  If the black-body fit presented in Bade et
al. (1996) is a good representation of the spectrum of the emission,
then the bolometric correction to the X-ray flux is not very large and
the observed light curve gives a meaningful measure of the energetics
and luminosity of the flare.  We note that, during and after the X-ray
outburst, the optical brightness did not show any significant
variation \citep{kom99}.  This confirms that the bulk of the fallback
emission probably did occur in X-rays.  (If the flare had substantial
hidden emission in the far ultraviolet band, it is hard to see how
there could have been no variations in the optical.)

In addition to fallback, a second independent stage of evolution is
expected when the circularized fallback material viscously accretes
onto the black hole. This stage is expected to develop much more
slowly and to be very much dimmer than the fallback stage (Appendix
A).  It is not relevant for the particular flare in NGC~5905
considered here.

The time lag between the estimated disruption time $t_D=1990.36$ yr of
the event and the time $t_1$ corresponding to the peak of the light
curve gives the orbital period $\Delta t_1$ of the most tightly bound
debris.  The data constrain $\Delta t_1$ fairly tightly (eq.
[\ref{t1l}]).  By the arguments described in \S3, this translates into
constraints on the mass $M_H$ of the black hole and the mass $M_\star$
of the disrupted star.  The constraints are shown in Figure
\ref{fig2}.

\begin{figure}
\epsscale{0.83}
\plotone{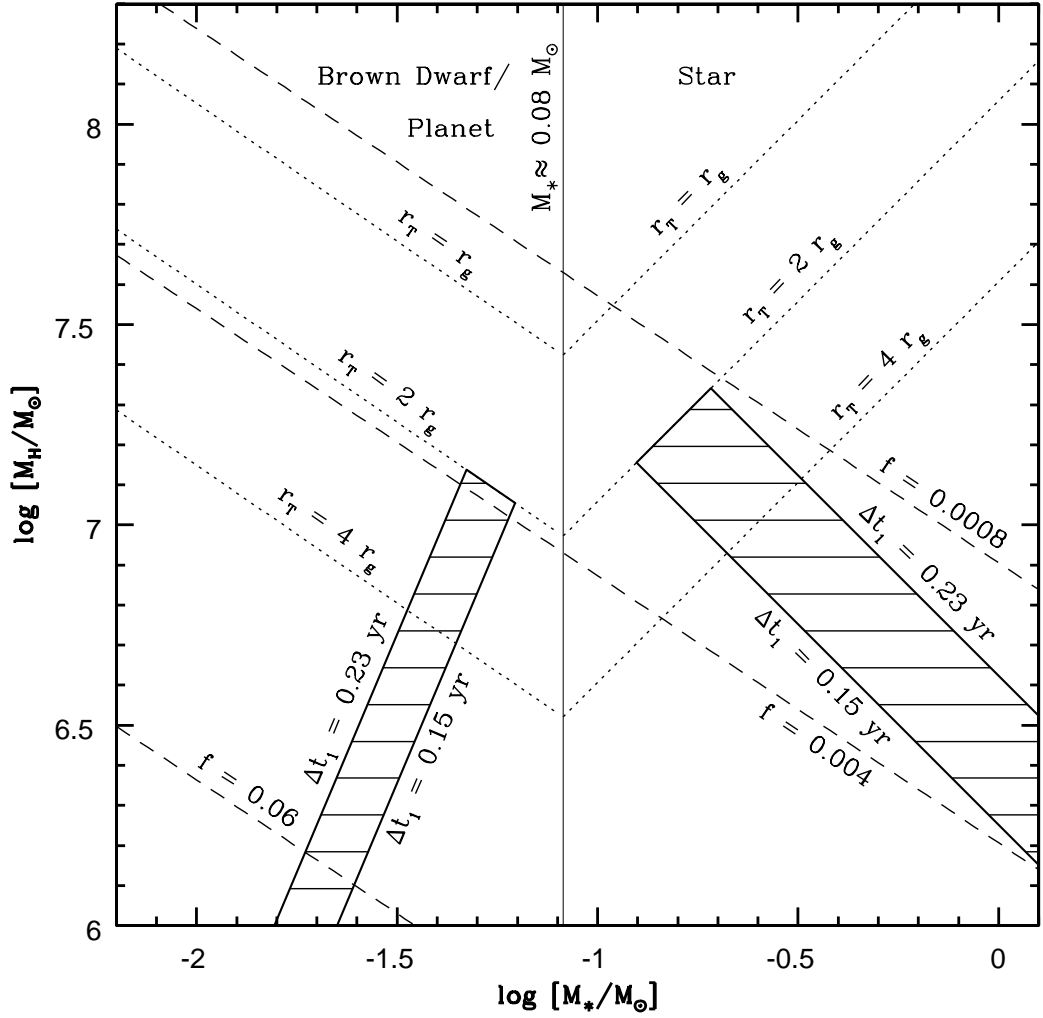}
\caption{Same as Fig.~\ref{fig2}, but assuming that the star experiences
negligible spin-up ($k=1$).
\label{fig3}}
\end{figure}

\begin{figure}
\epsscale{1}
\plotone{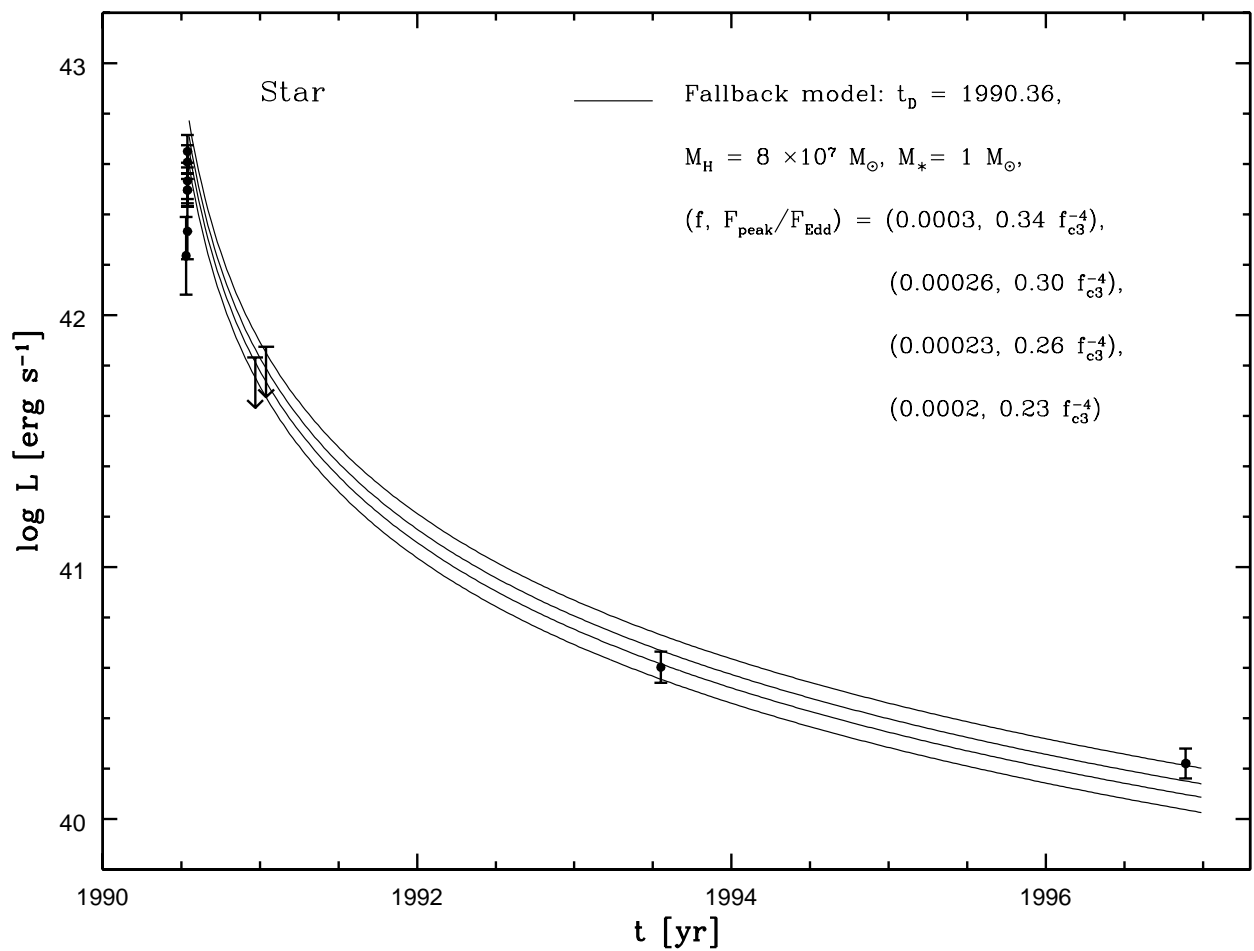}
\caption{Comparison of light curves predicted by the fallback model
with data: the disrupted object is assumed to be a low mass star.  The
parameter $f$ is the ratio of the accreted mass to the total mass of
the star, $F_{\rm peak}/F_{\rm Edd}$ is the ratio of the peak flux
$\equiv L_{\rm peak}/(\pi R_X^2)$ to the Eddington flux $\equiv L_{\rm
Edd}/(4 \pi r_T^2)$, where $f_{c3} \equiv f_c/3$.  Four models are
shown, each with the values of $f$ and $F_{\rm peak} /F_{\rm Edd}$
indicated (from top to bottom).
\label{fig4}}
\end{figure}

\begin{figure}
\epsscale{0.82}
\plotone{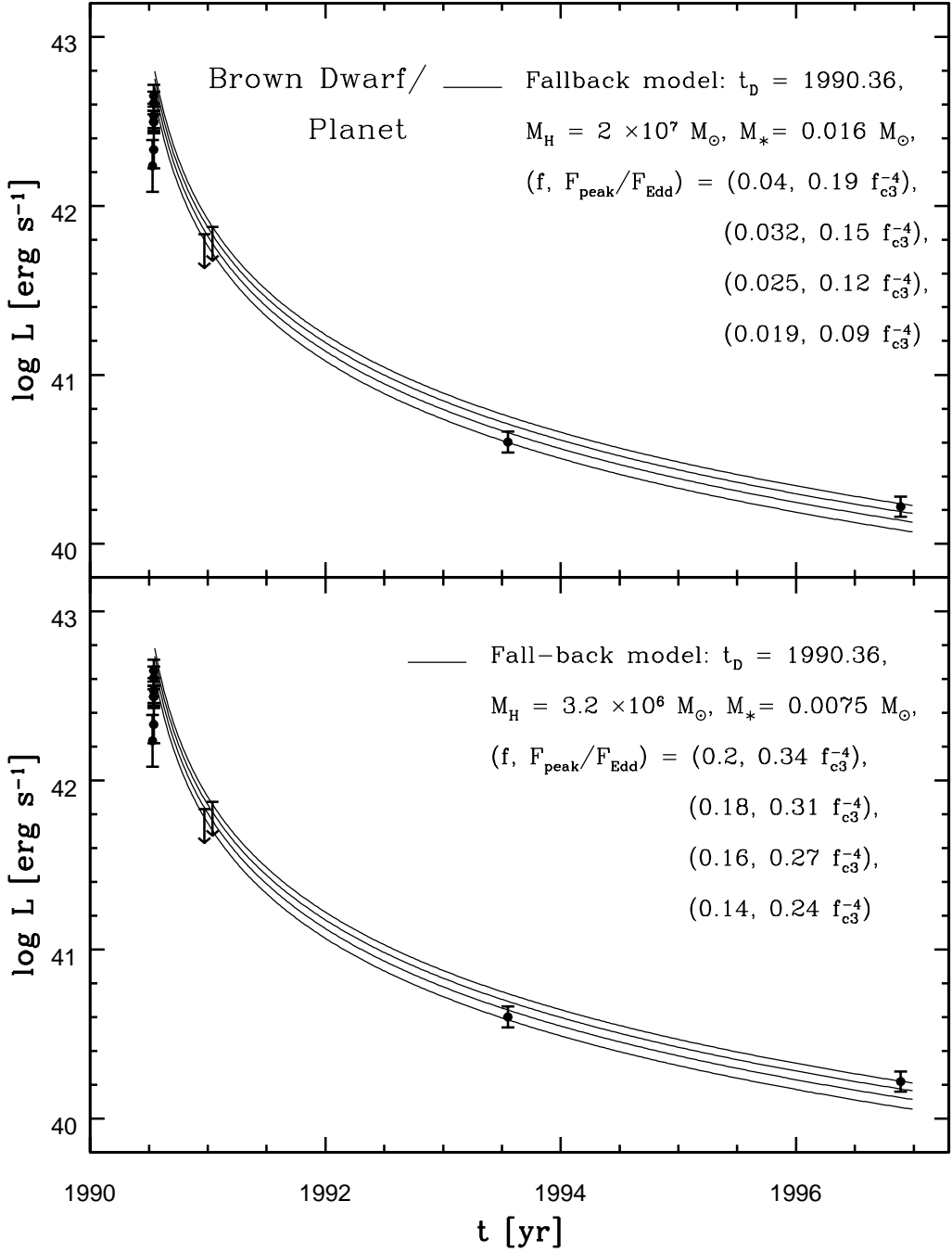}
\caption{Similar to Fig.~\ref{fig4}, but assuming that the disrupted object 
is a brown dwarf or a planet.  Two cases are shown:
$M_H = 2 \times 10^7 M_\odot$, brown dwarf of mass $M_\star = 0.016 M_\odot$
(upper panel); and $M_H = 3.2 \times 10^6 M_\odot$, planet of mass $M_\star 
= 0.0075 M_\odot$ (lower panel).
\label{fig5}}
\end{figure}

In deriving equation (\ref{t1}), we have estimated the binding energy
of the most tightly bound debris assuming that the star is fully spun
up by tidal interaction before being disrupted.  This corresponds to
a spin-up parameter $k=3$ (\S3); this value is supported both by
linear analysis \citep{pre77} and numerical simulations
\citep{ale01a}.  If on the other hand there is no tidal spin-up, then
we expect $k=1$, and the results change significantly
(Fig. \ref{fig3}).  We consider $k=1$ to be somewhat unlikely, and
have focused on the $k=3$ results.  

There is also ambiguity in the value of the parameter $\mu$ in the
definition of $r_T$ [eq. (\ref{rT})].  Numerical simulations show that
$\mu$ is usually of order unity, the value we use, but that it can be
as large as $1.7$ in specific cases \citep{fro94,die97}.  The time 
interval $\Delta t_1$
[eq. (\ref{t1})] is very sensitive to the value of $\mu$: $\Delta t_1
\propto \mu^3$.  Because of the tight observational constraint on
$\Delta t_1$ (Figs. \ref{fig2} and \ref{fig3}), $\mu$ cannot be much
larger than $1$ since otherwise it would imply too small a black hole
mass.

Another constraint on $M_H$ and $M_\star$ is obtained from the
requirement that the radius $r_T$ at which the star is tidally
disrupted must lie outside the radius of the marginally bound circular
orbit.  This requires $r_T>4r_g$ for a Schwarzschild black hole and
$r_T>r_g$ for the most favorably oriented orbit around a maximally
spinning Kerr hole.  The uncertainty in the limiting $r_T$ is
indicated by the dotted lines in Figure \ref{fig2}.  For concreteness,
we have assumed that the limiting radius is intermediate between the
two extreme cases mentioned above; we have used the limit $r_T>2r_g$.
Yet another source of uncertainty derives from the fact that we have
used Newtonian dynamics in our analysis, whereas it is clear that
relativistic effects should be important 
\citep{lum85,lag93,fro94,die97}.

Ignoring these caveats, we obtain a tight correlation between the mass
of the disrupted star in NGC~5905 and the mass of the central black
hole.  As Figure \ref{fig2} shows, the data are consistent with two
distinct solution branches.

One solution branch corresponds to a relatively massive black hole,
$M_H$ close to $10^8M_\odot$, tidally disrupting a main sequence star
with $M_\star\sim M_\odot$.  The other branch consists of a lower mass
black hole, $M_H < 10^{7.5}M_\odot$, disrupting a brown dwarf or a
giant planet with $M_\star \la 0.02M_\odot$.  Assuming that NGC~5905
satisfies the well-known correlations between black hole mass and host
galaxy properties \citep{sal00,geb00,mer01}, the black hole mass is
estimated to be $M_H\sim 10^7-10^8M_\odot$ \citep{kom01}.  This mass
range permits both solution branches.

Figures \ref{fig4} and \ref{fig5} show sample theoretical light curves
corresponding to the two solution branches.  In each case, there is
one free parameter, namely the fraction $f$ of the mass of the
disrupted star that returns as fallback debris to produce the observed
radiation.  We find that $f$ is unusually small if the disrupted star
is a main sequence star; for the particular example shown in Figure
\ref{fig4}, we require $f<0.0003$, which is an extremely small value
compared to the theoretically expected value of $f\sim 0.1-0.5$.
Brown dwarf models give somewhat larger estimates for $f$
(Fig.~\ref{fig5}), though still uncomfortably small, unless the black
hole has an unusually low mass ($<10^{6.5}M_\odot$). Though the
estimated $f$ can be increased by adopting a larger ambiguity factor
$\mu$ in the definition of $r_T$, we do not consider this likely since
a larger $\mu$ will lead to an unacceptably small value of $M_H.$

The above results on $f$ are not surprising.  As we showed in \S2, the
total radiative energy in the flare is only $4\times10^{49}$ erg,
which for any reasonable radiative efficiency corresponds to an
unexpectedly small mass $\sim {\rm few}\times10^{-4}M_\odot$.  Figures
\ref{fig4} and \ref{fig5} merely confirm this result using a detailed
estimate of the radiative efficiency (eq. \ref{ep}).

Can $f$ be so small for a tidally disrupted solar mass star?
\citet{ree88} and \citet{ren95} have argued that a star could lose
just its outer layers if the minimum distance from the star to the
black hole is somewhat larger than $r_T$. Here we point out a related
possibility.  The usual formula for the tidal disruption radius, $r_T
= R_\star (M_H / M_\star)^{1/3}$, is correct only for a homogeneous
star with a uniform mass density. For a real star, we could define a
variable tidal disruption radius,
\begin{eqnarray}
        r_T(R) = R \left[\frac{M_H}{M(R)}\right]^{1/3}
               = \left[\frac{3 M_H}{4\pi \overline{\rho}(R)}\right]^{1/3}\,,
\end{eqnarray}
where $R \le R_\star$ is a radius inside the star, $M(R)$ is the mass
contained within $R$, and $\overline{\rho} (R)$ is the corresponding
mean density. The meaning of $r_T(R)$ is as follows: if the star is
located at a distance $r = r_T(R)$ from the black hole, then the part
of the star outside radius $R$ (i.e., the spherical envelope between
$R$ and $R_\star$) is tidally stripped off, and the core inside $R$
remains intact.  Obviously, $r_{T,\max} = r_T(R_\star)$ and
$r_{T,\min} = (3 M_H / 4 \pi \rho_c)^{1/3}$, where $\rho_c$ is the
central density of the star.  If the star is at a distance $r$ from
the black hole such that $r_{T,\min} < r < r_{T,\max}$, then only a
part of the envelope of the star is peeled off.

One important question is how the core of the star responds when it
loses its envelope.  What is relevant is the response on a dynamical
time.  Webbink (1985) has calculated the logarithmic derivative of the
radius of a main sequence star with respect to mass,
$\zeta_s\equiv(\partial\ln R_\star/\partial\ln M_\star)_s$, where the
subscript $s$ indicates that the entropy profile of the star is held
fixed (as appropriate for a dynamical process).  Consider a star that
loses a fraction of its mass as a result of a tidal interaction and
assume that it dynamically adjusts to a new equilibrium corresponding
to its reduced mass.  If $\zeta_s>1/3$, then the mean density of the
new configuration will be higher than the original mean density of the
star.  The leftover mass is then stable to further stripping.
However, if $\zeta_s<1/3$, the new configuration will be prone to
further tidal mass loss, and one has a runaway situation in which the
entire star is likely to be disrupted.

Webbink (1985) shows that $\zeta_s>1/3$ for $\log(M_\star/M_\odot) \ga
-0.2$.  Since this is precisely the mass range of interest if the
disrupted star in NGC~5905 was a main sequence star (Fig.~\ref{fig2}),
we conclude that partial stripping is a viable possibility.  To
estimate what fraction of its mass the star might lose, consider a
star like the Sun as an example.  For the Sun, the mass inside a
radius $R_1 = 0.7 R_\odot$ is $0.975 M_\odot$, and correspondingly,
$r_T(R_1) = 0.706\, r_{T,\max}$.  Thus, for a range of distances between
$r_{T,\max}$ and $0.7\, r_{T,\max}$ (a non-negligible range), such a 
star will lose only a couple of percent of its mass.  If we assume that
10\% of this mass returns via fallback, then we can have $f$ as small
as $\sim0.002$.  This is still much larger than the value needed to
explain the flare in NGC~5905, but it is at least within striking
distance.

Fully convective lower main sequence stars and degenerate brown dwarfs
have $\zeta_s\sim-1/3$.  These stars are likely to be fully disrupted
in tidal encounters since they expand to a {\it larger} radius as they
lose mass.

Apart from the small value of $f$, another unexpected result of our
analysis is that the peak radiative flux emitted during the flare,
$F_{\rm peak}=\sigma T_{\rm eff}^4=\sigma (T_{\rm bb}/f_c)^4$, is
close to the Eddington flux.  (Here $T_{\rm bb}$ is the black-body
temperature of the radiation and $f_c$ is a color correction factor.)
The results are shown for our models in Figures \ref{fig4} and
\ref{fig5}.  Even with $f_c\sim 3$ (\S2), we find $F_{\rm peak}/F_{\rm
Edd}\sim 0.1-0.3$; the value increases rapidly with decreasing $f_c$
(as $f_c^{-4}$).  What makes this surprising is that the {\it
luminosity} of the source, even at the peak of the flare, is much
below the Eddington luminosity: compare $L_{\rm peak}\sim
4.5\times10^{42} ~{\rm erg\,s^{-1}}$ with $L_{\rm Edd}\sim
10^{45}(M_H/10^7 M_\odot) ~{\rm erg\,s^{-1}}$.  Clearly, the radiation
in the flare must have been emitted from a small patch on the source,
and the patch must have projected a small solid angle at the central
black hole.  This result was anticipated in the model-independent
analysis of \S2, where we showed that the effective size of the
emitting patch is small compared to the area we might expect if the
entire accretion disk out to a few $r_g$ is involved in the emission.

The small size of the emitting region can be understood in analogy
with mass transfer binary stars.  In the latter systems, it is known
that the cold gas stream from the secondary star impacts the accretion
disk at a compact ``bright spot'' on the outer edge of the disk.  Most
of the radial kinetic energy of the stream is converted to thermal
energy near this spot and radiated locally.  The analogy with the
tidal disruption problem is close because the returning material after
a disruption event comes in similarly as a narrow cold stream
\citep{koc94,kim99}.  It is therefore natural that the radiation
during the fallback stage should be emitted from a relatively small
patch on the source.  The effect is probably even more severe in the
tidal disruption problem.  Lense-Thirring precession may well cause
the incoming stream to be misaligned with respect to the orbital plane
of the accretion disk \citep{koc94}, causing the geometric area of the
impact spot to become smaller.  (We note that in binaries the stream
appears to skim over the surface of the disk after impacting at the
bright spot, thereby increasing the effective area over which the
kinetic energy of the stream is dissipated.  This effect would be
reduced if the orbit of the stream is not in the plane of the disk.)

Having decided that emission from a small patch is natural for this
problem, we should next ask whether it is a pure accident that the
flux from the patch in the NGC~5905 flare happens to be nearly equal
to the Eddington flux.  We would like to suggest that it is perhaps
not an accident.

Imagine a narrow stream with a mass accretion rate $\dot M$ impacting
on the accretion torus at a radius $r_T$ and over a small patch that
subtends a solid angle $\Delta \Omega_{\rm stream}$ at the black hole.
Let us suppose that the kinetic energy in the stream is thermalized
and radiated immediately from the patch.  The radiated flux is locally
given by $\epsilon \dot M c^2/\Delta\Omega_{\rm stream}r_T^2$.  What would
happen if this flux is greater than the local $F_{\rm Edd}$?  Jeremy Goodman
(private communication) suggests that the super-Eddington flux would
cause the local radiating atmosphere to puff up and that the radiation
would be emitted from a larger patch than the original impact region.
Moreover, it would be natural for the radiating patch to self-adjust
so that the flux is close to the Eddington limit.  According to this
picture then, the radius $R_X$ estimated in equation (\ref{rx})
refers to the size of the puffed up atmosphere, while the actual
transverse size of the stream is smaller.  If the stream is extremely
narrow, then the flux would be close to Eddington during a significant
fraction of the light curve.  Equivalently, the black-body temperature
of the radiation would remain constant, even while the luminosity
decreases.

This picture (which is predicated on the assumption of a narrow
fallback stream) suggests that most tidal disruption events should be
visible in soft X-rays; the condition $F=F_{\rm Edd}$, coupled with
$f_c\sim3$ (see \S2), ensures that the blackbody temperature of the
radiation will be $\ga50$ eV for a $10^7M_\odot$ black hole, which is
well-matched to X-ray observations.  Soft X-ray surveys should thus be
well-suited for an investigation of both the frequency of tidal
disruption events and the amount of mass accreted in each event. In
this connection, it is interesting that, with the exception of the UV
flare of Renzini et al. (1995), all known candidate flares for tidal
disruption events have been identified in the X-ray band (see Komossa
2001 for five candidates and Choi et al. 2002 for a possible sixth
candidate).

Menou \& Quataert (2001) have emphasized that, given current estimates
of the rate of tidal disruptions, the nuclei of many nearby normal
galaxies should be much brighter than observed if a fair fraction of a
stellar mass is accreted by the black hole after each tidal disruption
event.  This is not relevant for the late accretion phase of the
particular NGC~5905 flare discussed here because, as we have discussed
(see also the Appendix below), very little mass ($\ll 1 M_\odot$)
returned to the BH in this case.  However, whether we attribute the
small amount of mass in the NGC~5905 flare to partial stripping of a
star or to the disruption of a substellar object, it does not help to
explain the general problem of why nearby galactic nuclei are so
faint.  In addition to partial disruptions of the sort that might have
occurred in NGC~5905, we expect total stellar disruptions in
significant numbers and these should provide a significant mean mass
accretion rate in galactic nuclei.  Disruptions of substellar objects
only add to the purely stellar disruptions considered to date.  Thus,
alternative ways of explaining the dimness of nearby galactic nuclei
are required (Menou \& Quataert 2001; Narayan 2002), beyond the
specific scenario outlined here for the giant flare of NGC~5905.
Either the mass must accrete in a radiatively inefficient mode or the
accretion must be suppressed or slowed down significantly.

\acknowledgments 

The authors gratefully thank Stefanie Komossa for kindly providing the
{\it ROSAT} data listed in Table 1 along with the conversion factors
from count rate to X-ray luminosity.  The authors also thank Jeremy
Goodman, Bohdan Paczy\'nski and the referee for helpful comments. LXL
and RN thank the Institute for Advanced Study and the Department of
Astrophysical Sciences, Princeton, for hospitality while this work was
being done. LXL's and KM's research was supported by NASA through
Chandra Postdoctoral Fellowship grant numbers PF1-20018 (LXL) and
PF9-10006 (KM) awarded by the Chandra X-ray Center, which is operated
by the Smithsonian Astrophysical Observatory for NASA under contract
NAS8-39073.  RN's research was supported in part by NSF grant
AST-9820686.

\appendix
\section{Viscous Evolution of the Thin Disk Formed from Tidal Disruption}
\label{appa}

After circularization, the fallback material from a tidal disruption
event will form a torus at a radius $r\sim2r_T$.  The torus will
then spread out viscously and accrete onto the black hole.

The time-scale for this stage of the evolution is determined by the
viscous time-scale \citep{can90,ulm99}
\begin{eqnarray}
        t_{\rm vis} = \frac{t_{\rm Kep}(2 r_T)}{\pi\alpha h^2}
                    \approx 2.9 \times 10^3\, {\rm yr}
                    \left(\frac{\alpha}{0.1}\right)^{-1}
                    \left(\frac{h}{10^{-3}}\right)^{-2} 
                    \left(\frac{M_\star}{M_\odot}\right)^{-1/2}
                    \left(\frac{R_\star}{R_\odot}\right)^{3/2}\,,
        \label{tvis}
\end{eqnarray}
where $t_{\rm Kep}$ is the Keplerian orbital period, $\alpha$ is the
standard viscosity parameter \citep{sha73}, and $h$ is the ratio of
the disk height to radius.  It is easy to check that for the case of
the flare in NGC~5905, where $L_{\rm peak} \sim 10^{-3} L_{\rm Edd} 
(M_H/10^7 M_\odot)^{-1}$, the accretion disk would correspond to the
regime of the ``middle disk'' in which electron scattering opacity
dominates over the Kramers opacity and gas pressure dominates over
radiation pressure \citep{sha73,nov73,fkr92}. Then, $h$ is estimated
to be
\begin{eqnarray}
        h \approx  1 \times 10^{-3}\, 
            \left(\frac{\alpha}{0.1}\right)^{-1/10}
            \left(\frac{\dot{M}_2}{10^{-3} \dot{M}_{\rm 
            Edd}}\right)^{1/5}
            \left(\frac{M_H}{10^7 M_\odot}\right)^{-1/10}
            \left(\frac{r}{r_H}\right)^{1/20}\,,
        \label{h1}
\end{eqnarray}
where $\dot{M}_2$ is the mass accretion rate, $\dot{M}_{\rm Edd} \equiv
L_{\rm Edd}/(0.1 c^2)$, and $r$ is the radius in the disk. In equation
(\ref{h1}) we have neglected a dimensionless factor $\sim 1$.  

Inserting equation (\ref{h1}) into equation (\ref{tvis}) and setting
$r = 2 r_T$, we obtain
\begin{eqnarray}
        t_{\rm vis} &\approx& 2.3 \times 10^3\,{\rm yr}\, 
            \left(\frac{\alpha}{0.1}\right)^{-4/5}
            \left(\frac{\dot{M}_2}{10^{-3} \dot{M}_{\rm 
            Edd}}\right)^{-2/5} \nonumber\\
            &&\times\left(\frac{M_H}{10^7 M_\odot}\right)^{4/15}       
            \left(\frac{M_\star}{M_\odot}\right)^{-7/15}
            \left(\frac{R_\star}{R_\odot}\right)^{7/5}\,.
        \label{tvis1}
\end{eqnarray}
The large value of $t_{\rm vis}$ confirms that the flare in NGC~5905 
corresponded to the fallback stage, not the viscous accretion stage.

From the discussion in \S\ref{sec3} on the fallback and viscous accretion
stages associated with a tidal disruption event, the ratio of the total
energy radiated during the accretion stage to the total energy radiated
during the fallback stage is $2 r_P / r_{ms} - 1$. Therefore, we can estimate
the ratio of the peak of the viscous luminosity to the peak of the fallback
luminosity by
\begin{eqnarray}
        \frac{L_{\rm vis}}{L_{\rm fb}} &\approx& \left(\frac{2 r_P}
                {r_{ms}}- 1 \right)\frac{\Delta t_1}{t_{\rm vis}}
                \nonumber\\
                &\approx&
                2.4 \times 10^{-5}\, \left(\frac{\alpha}{0.1}\right)
                \left(\frac{h}{10^{-3}}\right)^2
                \left(\frac{M_H}{10^7 M_\odot}\right)^{1/2}
                \left(\frac{M_\star}{M_\odot}\right)^{-1/2}
                \left(\frac{2 r_P}{r_{ms}}- 1 \right)\,,
        \label{ratio}
\end{eqnarray}  
where equations (\ref{t1}) and (\ref{tvis}) have been used.

The extremely large value of $t_{\rm vis}$ and the extremely small value 
of $L_{\rm vis}/L_{\rm fb}$ imply that the accretion stage will never be 
observed in practice for the particular case of NGC~5905. [In fact, 
according the above estimates, one may expect a new tidal disruption 
before the disk evolves significantly \citep{sye99,mag99}.]


\clearpage
\begin{deluxetable}{lllll}
\tablewidth{0pt}
\tablecaption{{\it ROSAT} Data on the NGC~5905 Flare \label{tab1}}
\tablehead{
\colhead{TIME\tablenotemark{a}} & \colhead{CR\tablenotemark{b}} & 
\colhead{ERR\_CR\tablenotemark{c}} & \colhead{LUMX\tablenotemark{d}} & 
\colhead{ERR\_LUMX\tablenotemark{e}}
}
\startdata
1990.53 & 2.2981E$-$1 & 8.1771E$-$2 & 1.72      & 0.61     \\
1990.54 & 2.8713E$-$1 & 7.3284E$-$2 & 2.15      & 0.55     \\
1990.54 & 4.1855E$-$1 & 6.2263E$-$2 & 3.14      & 0.47     \\
1990.54 & 4.5519E$-$1 & 7.5933E$-$2 & 3.41      & 0.57     \\
1990.54 & 4.1786E$-$1 & 6.2403E$-$2 & 3.13      & 0.47     \\
1990.54 & 5.9612E$-$1 & 8.7399E$-$2 & 4.47      & 0.66     \\
1990.54 & 5.4012E$-$1 & 8.2004E$-$2 & 4.05      & 0.62     \\
1990.97 & $<$0.09     & ------      & $<$0.68   & ------   \\
1991.04 & $<$0.10     & ------      & $<$0.75   & ------   \\
1993.55 & 7.0000E$-$3 & 1.0000E$-$3 & 4.00E$-$2 & 5.71E$-$3\\
1996.89 & 2.9000E$-$3 & 4.0000E$-$4 & 1.66E$-$2 & 2.29E$-$3\\
\enddata

\tablenotetext{a}{Time of measurement in years.}
\tablenotetext{b}{Count rate in cts s$^{-1}$.}
\tablenotetext{c}{Error in the count rate.}
\tablenotetext{d}{X-ray luminosity in units of $10^{42}$ erg~s$^{-1}$.}
\tablenotetext{e}{Error in the X-ray luminosity.}
\tablecomments{The data were provided by S. Komossa. The conversion 
from count rate to luminosity was done
assuming a black-body spectrum and a Hubble
constant of $H_0 = 50$ km s$^{-1}$ Mpc$^{-1}$.  For this model, 1 cts~s$^{-1}$  
corresponds to  $\sim 7.5\times 10^{42}$ erg~s$^{-1}$ during 1990 when the
source was in the high-state, and  $\sim 5.7
\times 10^{42}$ erg~s$^{-1}$ during 1993 and 1996 when the source had a flatter
spectrum (S. Komossa 2001, private communication).}

\end{deluxetable}

\clearpage

\clearpage

\clearpage

\clearpage

\clearpage

\end{document}